\documentclass[a4paper,fleqn,usenatbib]{mnras}
\usepackage{newtxtext,newtxmath}
\usepackage[T1]{fontenc}
\usepackage{ae,aecompl}

\usepackage{graphicx}
\usepackage{amssymb}
\usepackage{dcolumn}% Align table columns on decimal point
\usepackage{bm}% bold math

\usepackage{lineno}

\def\ba{\begin{array}}
\def\ea{\end{array}} 
\def\bea{\begin{eqnarray}}
\def\eea{\end{eqnarray}}
\def\beq{\begin{equation}}
\def\eeq{\end{equation}}
\def\ben{\begin{enumerate}}
\def\een{\end{enumerate}}

\def\brr{\begin{array}}
\def\err{\end{array}}

\def\calC{\S}

\title[The size of our causal Universe]{The size of our causal Universe}
\author[Enrique Gazta\~naga]{Enrique Gazta\~naga\thanks{E-mail:gaztanaga@gmail.com}
\\ \\
Institute of Space Sciences (ICE, CSIC), 08193 Barcelona, Spain \\
Institut d\'~Estudis Espacials de Catalunya (IEEC), 08034 Barcelona, Spain
}

\date{\today}

\begin{document}
\label{firstpage}
\pagerange{\pageref{firstpage}--\pageref{lastpage}}
\maketitle

\begin{abstract}
%% Text of abstract
A Universe with finite age also has a finite causal scale. Larger scales can not 
affect our local measurements or modeling, but far away locations could have different cosmological parameters. The size of our causal Universe depends on the details of inflation and is usually assumed to be larger than our observable Universe today.  To account for causality, we propose a new boundary condition, that can be fulfill by fixing  the cosmological constant  (a free geometric parameter of gravity). This forces a cancellation of vacuum energy with the cosmological constant.  As a consequence, the measured cosmic acceleration can not be explained by a simple cosmological constant or constant vacuum energy. We need some additional odd properties such as the existence of evolving dark energy (DE) with energy-density fine tuned to be twice that of dark matter today.
We show here that we can instead explain cosmic acceleration without DE (or modified gravity)  assuming that the causal scale is smaller than the observable Universe today. Such scale corresponds to half the sky at z=1 and 60 degrees at z=1100, which is consistent with  the anomalous lack of correlations  observed in the CMB.
%{\color{black} Late time cosmic acceleration can then be interpreted as the smoking gun of primordial Inflation.}

%but could be confirmed or falsified by further observations on very large scales.
\end{abstract}

%Cosmology, Inflation


%%
%% Start line numbering here if you want 
%%
%\linenumbers

%% main text
\section{Introduction}

\label{S:1}
One of the most striking changes to Newton's  gravity proposed by Einstein is that energy gravitates. 
%This resulted in the famous light deflection Eclipse experiment  on 1919.
Scientists have since been wondering if vacuum energy $\rho_{\rm vac}$ (vacuum fluctuations, zero-point fluctuations, quantum vacuum, dark energy or aether) could also gravitate. 
 Measurements of cosmic acceleration (see e.g. \citealt{Planck2018,des2018,Tutusaus2017})  point to a model with $\Lambda$, that we refer to as $\Lambda$CDM and could be interpreted as a consequence of the gravity of  $\rho_{\rm vac}$.
Even while the accuracy and precision of  measurements have greatly improved in the last years,
the  mean values of cosmological parameters have remained similar for well over a decade (see e.g. \citealt{Gaztanaga2006,Gaztanaga2009}). 
The Friedmann-Lemaitre-Robertson-Walker (FLRW) flat metric in comoving coordinates $(t,\chi)$:
\beq
ds^2= g_{\mu\nu} dx^\mu dx^\nu = dt^2 - a(t)^2( d\chi^2 + \chi^2 d\Omega^2 )
\label{eq:frw}
\eeq
is the exact general solution for a mathematically homogeneous and isotropic flat Universe.
The scale factor, $a(t)$, describes the expansion of the Universe as a function of time. We can relate $a(t)$ to the energy content of the Universe for a perfect fluid by solving the field equations (see Eq.\ref{eq:rmunu}-\ref{eq:Tmunu}):
\bea
R_0^0= R_{00} &=& - \left( 
\frac{3 \ddot a}{a}\right) =  4\pi G (\rho + 3p)- \Lambda
\label{eq:Hubble0}
\\ \label{eq:Hubble}
H^2 &\equiv& \left({\dot{a}\over{a}}
\right)^2 
=  {8\pi G\over{3}} \rho +  {\Lambda\over{3}}
\eea
where $\rho = \rho_m a^{-3}+\rho_r a^{-4} +\rho_{\rm vac}$
and $\rho_m$ is the pressureless matter density today ($a=1$),
$\rho_r$ corresponds to radiation (with pressure $p_r=\rho_r /3$)  and $\rho_{\rm vac}$  represents vacuum energy ($p_{vac}= -\rho_{\rm vac} $).
\footnote{ {\color{black} For easy of notation we focus on the flat case, but our results can easily be extended to the non-flat case or non trivial topology.} }
 One can argue that 
$\Lambda$ is indistinguishable from $\rho_{vac}$, because
field equations are degenerate to the combination:
\beq
\rho_{\Lambda}  \equiv  \rho_{\rm vac} + \frac{\Lambda}{8\pi G}  .
\label{eq:rhoHlambda}
\eeq
Here we take $\Lambda$ to be a fundamental (geometrical) constant, while $\rho_{vac}$ depends on the actual energy content of our universe. The measured $\rho_\Lambda$ is very small compared to what we expect for $\rho_{\rm vac}$. 
Moreover, $\rho_\Lambda \simeq 2.3\rho_m$ today, which seems a remarkable coincidence. 
Possible solutions to this puzzle are:  I) $\Lambda=0$ and
 $\rho_\Lambda$ originates only from $\rho_{\rm vac}$ or some dark energy (DE) \citep{Weinberg1989,Elizalde1990, Carroll,Huterer,Elizalde2006}, II)  $\rho_{\rm vac}=0$ and we need to fix $\Lambda$ or Modified Gravity \citep{Lobo2001,Gaztanaga2002,Lue,Nojiri2017} or III) there is a cancellation between $\Lambda$ and $\rho_{\rm vac}$,  as we will propose here.

Eq.\ref{eq:frw}-\ref{eq:Hubble} are a mathematical extrapolation. 
A physical model requires a mechanism to produce homogeneity that respects causality. 
Note that the Lorentz invariance or covariance of the field equations in General Relativity (GR) is not enough to warrant that a given solution has a causal structure (see \citealt{Minguzzi,Howard} and references therein).  This is not the case for the  FLRW model, because $\rho$ and $p$ are the same everywhere at any fix cosmic time, and this can not have causal explanation for a Universe that has a finite age.
If we want a causal explanation for our Universe we can not find the solution directly solving the Cauchy problem in GR, because we are not allowed to setup  acausal initial conditions for such a problem. The only way to do this is to setup initial conditions that are random with no correlations. This is what happens in the realm of vacuum quantum fluctuations under Heisenberg uncertainty, which do not require a causal mechanism to exist.
Inflation could them produce a large and homogeneous universe out of these initial quantum fluctuations. But even for such a case, there is a finite causal scale associated with the duration of inflation.

\begin{figure}
\centering \includegraphics[width=\linewidth]{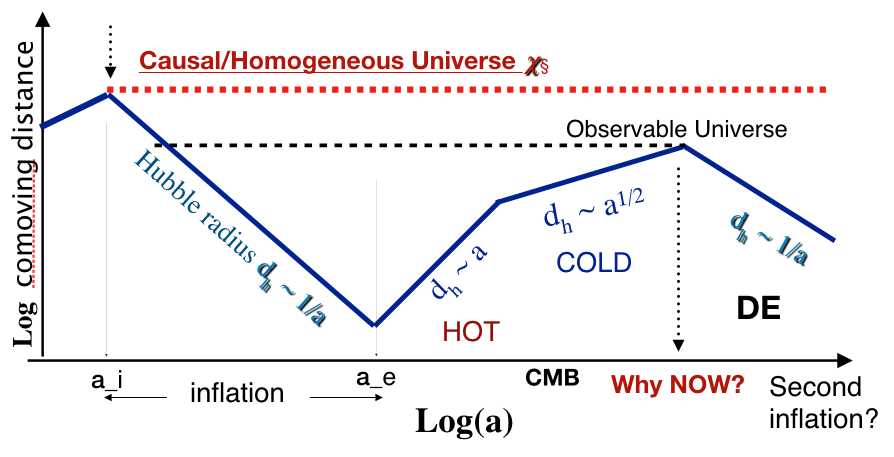}
\caption{Comoving 
Hubble radius  $d_H = c/(aH) $ (blue line) as a function of the scale factor $a$.  The Causal Universe $\chi_\S$ is identified with the region inside the largest causally connected scale at the beginning of inflation (red dashed line). A mysterious Dark Energy (DE) produces a second inflation around our time (label "now" in the figure) that makes $d_H$ decrease again.}
\label{Fig:horizon0}
\end{figure}

Particles separated by distances larger than the comoving Hubble radius 
$d_H(t)=c/[a(t)H(t)]$ can't communicate at time $t$. Distances larger than the horizon
\beq
\eta(a) = c \int_{0}^{t}  \frac{dt}{a(t)} = \int_{0}^{a}  {d\ln(a)} ~d_H(a) ,
\label{eq:eta}
\eeq
have never communicated.
We know from the cosmic microwave background (CMB) and large scale structure (LSS) that the Universe was very homogeneous on scales that were not causally connected (without inflation). 
This either means that the initial conditions where acausally  smooth to start with or that there is a mechanism like inflation \citep{Dodelson,Liddle1999,Brandenberger} which inflates regions outside the Hubble radius.
During inflation, $d_H$  decreases  which freezes out communication on comoving scales larger than the horizon  $\chi_\calC \simeq \eta(a_i)=d_H(t_i)$ when inflation begins, at $a_i=a(t_i)$. Inflation also smoothed and stretches out any initial (quatum) inhomogeneities to scales that could be larger than our current horizon. This creates homogeneous and flat patches.
When inflation ends, radiation from reheating makes $d_H$ grow again. Thus, the scale $\chi_\calC$ is fixed before inflation in comoving coordinates and is the same for all times, while the horizon $\eta$ and $d_H$ change with time.
This is illustrated in Fig.\ref{Fig:horizon0}.
Inflation  allows the full observable Universe to originate from a very small causally connected  homogeneous patch, $\chi_\calC$, which could be as small as the Planck scale.  
We usually assume that this region $\chi_\calC$ is much larger that our observable universe today: $\simeq 3 c/H_0$.
As we approach our epoch (label "now" in the figure), we  believe that a mysterious Dark Energy (DE) produces a second inflation that makes $d_H$ decrease again.
{\color{black} Why a second inflation now? Are both inflations related?}

Regardless of the details of inflation,
a Universe of finite age will only be causally homogeneous 
for scales smaller than some cut-off $\chi<\chi_\calC$.
We need a boundary condition at $\chi=\chi_\calC$ to account for the lack of causality (and therefore homogeneity) at larger scales.  This results in  a cancellation between $\Lambda$ and $\rho_{vac}$ and could be the caused of the 
current cosmic acceleration. In \S2  we view this problem in Classical Physics, while in \S3 we present a relativistic version. In \S4 we estimate the size of the causal Universe and discuss the implications for inflation and CMB. We end with some Discussion and Conclusions.

\section{Gauss's law}
\label{sec:Hooke}

A key property of Gravity is Gauss's law.
The acceleration $\vec{\mathrm{g}}$ created by a point mass  at distance $\vec{\mathrm{r}}$  is such
that a spherical shell of arbitrary radial density $\rho(r)$
produces a field which is identical to a point source of equal mass  $m$ in its center. This condition alone can be used to define gravity in Classical Mechanics, but the solution  is more general than Newton's law (see Appendix \ref{sec:hook}):
\bea
\vec{\mathrm{g}} \equiv  -\vec{\nabla} \phi &=& -\left( \frac{G m}{r^3} - \frac{\Lambda}{3} \right) \vec{r}
\nonumber \\
\nabla^2 \phi &=& 4\pi G \rho_m -\Lambda .
\label{eq:hooke}
\eea
Using Stokes Theorem, the gravitational field  produces a flux around 2D closed surface $\partial V$:
\beq
\Phi= 
\oint_{\partial V} ~d\vec{\mathrm{r}} ~ \vec{\mathrm{g}}  = 
- \int_V  dV \left( 4\pi G  ~\rho_m  - \Lambda  \right) 
%=  - 4\pi G ~m +  \Lambda V
\label{eq:flux00}
\eeq
so the flux only depends on the total mass (and $\Lambda$) inside the boundary $\partial V$.
Note the second term with $\Lambda$ in  Eq.\ref{eq:hooke}-\ref{eq:flux00} which corresponds to Hooke's law, i.e. proportional to distance. These of course are the same equations that come from GR in the Newtonian limit (see below). One can then argue that $\Lambda$ is just part of the laws of
gravity, as it is allowed by the symmetries of both GR and Classical
Gravity.  Current cosmological observations clearly indicate that
$\Lambda \neq 0$. 
There is therefore no need for a search for 
some more exotic 'Dark Energy'' explanation. For this to be confirmed it would good if
this interpretation can be used to predict the actual value measured for the cosmic acceleration, which is something that we do in section \ref{sec:CMB}.

Physicist assume that particles should be free at infinity, because of lack of causality. This is  why boundary terms are usually neglected at infinity. In the same spirit,
 we will require here that test  particles should be free  ($\vec{\mathrm{g}}=0$) or more relevant for a fluid: that boundary terms should be zero (i.e. the flux $\Phi =0$), when outside causal contact, $r> r_\calC$.
For $ r_\calC \Rightarrow \infty$ this condition 
requires $\Lambda \Rightarrow0$, as otherwise 
$\vec{\mathrm{g}}$ and $\Phi$ diverge.
Observational evidence that $\Lambda \neq 0$ may then indicate that $r_\calC$ is finite. 
This agrees with the finite age of the Universe.
 From Eq.\ref{eq:flux00} the boundary condition $\Phi(r>r_\calC)=0$, implies:
 \beq 
  \Lambda= 4\pi G \rho_m(r<r_\calC)
\label{eq:lambda0}
\eeq 
 which is clearly related to the coincidence problem: $\rho_\Lambda \sim 2 \rho_m$.
 Lets next explore this same argument in GR. For this we first need to see what is the relativistic version of Eq.\ref{eq:hooke}-\ref{eq:flux00}.

%%%%%%%%%%%%%%%%%%%%%%%%%%%%%%%%%%%%%%%%%%
%%%%%%%%%%%%%%%%%%%%%%%%%%%%%%%%%%%%%%%%%%
\section{Relativistic case}
\label{sec:infinite}

The symmetries of Einstein's  field equations allow for a cosmological constant $\Lambda$ term %(\citealt{Landau1971,Weinberg1972}):
(\citealt{Landau1971}):\footnote{We use the sign conventions in \cite{Landau1971}, but use greek letters for 4D space-time indexes and latin for 3D spatial indexes.}
\beq
R_\mu^\nu + \Lambda \delta_\mu^\nu = 8\pi~G~(T_\mu^\nu-\frac{1}{2} \delta_\mu^\nu T) ,
%R_{\mu\nu} - {1\over{2}} R~g_{\mu\nu} - \Lambda g_{\mu\nu} = {8\pi~G~\over{c^4}}~T_{\mu\nu}.
\label{eq:rmunu}
\eeq
For a  perfect  fluid with  density $\bar{\rho}$ and pressure  $\bar{p} \equiv \omega \bar{\rho}$:

\beq
T_\mu^\nu =  (\bar{\rho}+\bar{p}) u_\mu u^\nu - \bar{p} \delta_\mu^\nu  
\label{eq:Tmunu}
\eeq
where both $\bar{p}$ and $\bar{\rho}$ could change with space-time. 
 For events comoving with the fluid we have $u^i=0$  so that  $u^\mu u_\mu = u^0 u_0=1$ , so that the time-time component of the Ricci curvature is:
%\beq
%R = -8\pi G \, T -4\Lambda = 8\pi   (\bar{\rho}-3\bar{p}) - 4\Lambda
%\label{eq:scalar}
%\eeq
\beq
R_0^0 = 4\pi G  (\bar{\rho}+3\bar{p}) - \Lambda 
\label{eq:Ricci00} \\
%R_1^1 = R_2^2 = R_3^3 &=& 4\pi   (\bar{p}-\bar{\rho}) - \Lambda
%\label{eq:Ricci11}
\eeq

\subsection{The generalised Gauss's law}

Consider perturbations around 
Minkowski metric $\eta_{\mu\nu} $ (i.e. around empty space):

\beq 
g_{\mu\nu} = \eta_{\mu\nu} + h_{\mu\nu}
\label{eq:h}
\eeq 
 where $h_{\mu\nu}$ are small corrections. 
 To linear order in  $h_{\mu\nu}$ 
 %(i.e. equation above Eq.102.8 in  \cite{Landau1971}) 
 we have $R_{00} = 1/2 ~ \square h_{00}$ \citep{Landau1971},
so that wee can defined the gravitational potential as  $\phi \equiv  h_{00}/2$ to find:

\beq
R_{00}  = R_0^0 =   \square \phi \equiv 
- \nabla_\mu \nabla^\mu \phi  = - \nabla_\mu {\mathrm{g}}^\mu 
\label{eq:R00}
\eeq
where ${\mathrm{g}}^\mu \equiv \nabla^\mu \phi$ is the covariant gravitational acceleration.
%This relation turns out to be exact for the metric of the inhomogeneous Universe \citep{Gaztanaga2019}.
We can combine this relation with Eq.\ref{eq:Ricci00} to find:

\beq
\square \phi = 4\pi G  ~(\bar{\rho}+ 3 \bar{p})  - \Lambda
\label{eq:R00b}
\eeq
which is the relativistic generalization of Poisson equation Eq.\ref{eq:hooke}.
{\color{black}
Thus the relativistic version of Gauss law in Eq.\ref{eq:flux00} is then 
\beq
\Phi = 
\oint_{\partial M} ~dx_\mu ~ 
{\mathrm{g}}^\mu 
=
- \int_{M}  \sqrt{-g} ~~d^4x  % R_{00}
~\left[  4\pi G  ~(\bar{\rho}+ 3 \bar{p})  - \Lambda \right]
\label{poisson3}
\eeq
where $M$ is the 4D volume inside the 3D hipersurface $\partial M$.
Traditionally, we take such boundary terms to be zero at infinity: $\Phi(\infty)=0$
 (see Eq.4.7.8 in \citealt{Weinberg1972}).}

We can reach a similar expression without the weak field approximation by noticing that the equivalent of Poisson equation is the covariant time-time component of the Field Equations $R_0^0$ in  Eq.\ref{eq:Ricci00}. We can then identify the flux directly with:

\beq
\Phi =  - \int_{M}  \sqrt{-g} ~~d^4x ~ R_0^0 
\label{poisson4}
\eeq
which for a perfect fluid gives the same result as Eq.\ref{poisson3}.

\subsection{Causal Boundary condition}
\label{sec:boundary}

As mentioned in the Introduction,
scales larger that $\chi_\calC$  
can have no effect on the metric or the curvature
of the universe around us.  
{\color{black}
Mathematically, this appears as retarder Green functions ($\phi(\chi,t)=\phi(\chi-ct)$) as solutions to the wave equation Eq.\ref{eq:R00b} (or Eq.\ref{eq:Ricci00}) with appropriate boundary conditions (typically  $\chi_\calC \rightarrow \infty$). }
This could result in a non-homogeneous solution for the metric of the Universe on very large scales (see \citealt{Gaztanaga2019}).
An observer situated at the edge of our causal boundary will find a similar solution, but could measure different cosmological parameters, because she sees a different patch of the initial conditions.  There should be a 
 smooth background across disconnected regions with  an infrared cutoff in the spectrum of inhomogeneities for $\chi>\chi_\calC$.
Solutions in different regions could be matched as in  \cite{Sanghai-Clifton}.

We usually assume that particles should be free at infinity, because of lack of causality: if there is no cause there should not be any effect. This is  why boundary terms are usually set to zero at infinity. For example, to reproduce the field equations Eq.\ref{eq:rmunu} in GR, from an action principle we need to neglect these boundary terms (e.g. \citealt{Landau1971,Weinberg1972}).
 On scales $\chi<\chi_\calC$ we  have a homogeneous expanding Universe with $\bar{\rho}=\rho$. On larger scales we 
require boundary terms to vanish.
In particular we will require
$\Phi(\chi>\chi_\calC)=0$ in Eq.\ref{poisson4}, so that there is no flux (i.e. no effects of gravity) beyond the causal scale. This implies: 

\beq
\frac{\Lambda}{8\pi G} =   \frac{1}{2M_{\calC}} \int_{M_{\calC}}   \sqrt{-g} d^4x  ~  (\rho+ 3p) \equiv
 \frac{<\rho>_\calC + 3 <p>_\calC}{2},
\label{eq:rhoH}
\eeq
where $M_{\calC}$ is the volume inside the lightcone to the surface $\partial M_{\calC}$, where  $\chi=\chi_\calC$.
Note how this condition is similar to the one found by \cite{Lombriser} and 
the mechanism for sequestering vacuum energy (\citealt{Kaloper2016})
from requiring an additional minimization of the Einstein-Hilbert action.
Recall how here the scale $\chi_\calC$ is fixed in comoving coordinates  while the horizon $\eta$ and $d_H$ change with time (see Fig.\ref{Fig:horizon0}).
This is consistent with a constant value for $\Lambda$ in Eq.\ref{eq:rhoH}.

%%%%%%%%%%%%%%%%%%%%%%%%%%%%%%%%%%%%%%%%%%
\subsection{Vacuum Energy does not gravitate}
\label{sec:vac}
Inside $\chi<\chi_\calC$, we can use  Eq.\ref{eq:frw}-\ref{eq:Hubble} 
with $\rho= \rho_m+\rho_r+\rho_{\rm vac}$ and $p= \rho_r/3-\rho_{\rm vac}$, so that 
we can write Eq.\ref{eq:rhoH}  as:
\beq
\frac{\Lambda}{8\pi G} =
 \frac{<\rho_m>_\calC}{2} + <\rho_r>_\calC- \rho_{\rm vac}
\equiv \rho_{\calC} - \rho_{\rm vac} ,
\label{eq:m2}
\eeq
where $\rho_{\calC}$ is the mean matter and radiation contribution in the integral of Eq.\ref{eq:rhoH}. The values of $\rho_m$ and $\rho_r$ evolve with space-time, so that $\rho_{\calC}$ is the average contribution inside the volume $M_\calC$, while the vacuum density contribution is constant (by definition).  We can combine Eq.\ref{eq:rhoHlambda} with  Eq.\ref{eq:m2}:

\beq
\rho_{\Lambda}  \equiv \frac{\Lambda}{8\pi G}  + \rho_{\rm vac} =  \rho_{\calC} - \rho_{\rm vac} +   \rho_{\rm vac} = {\rho}_{\calC} ,
%{\rho_{m\calC}  + \rho_{\gamma\calC}\over{2}} - \rho_{\rm vac}(t_{\calC}) + \rho_{\rm vac}
\label{eq:rhoH2}
\eeq
which shows that vacuum energy cancels out and can not affect the observed value of $\rho_{\Lambda}$. 
In this respect we can conclude that vacuum does not gravitate.
This result is independent of the value of $\chi_\calC$ or the value $\rho_{\rm vac}$,
which could both be infinite (as it follows from Quantum Field Theory in the case of $\rho_{\rm vac}$).

%%%%%%%%%%%%%%%%%%%%%%%%%%%%%%%%%%%%%%%%%%%%%%%%%%
%%%%%%%%%%%%%%%%%%%%%%%%%%%%%%%%%%%%%%%%%%%%%%%%%%
\section{The size of our Causal Universe}
\label{sec:size}
Sometime in our past, at time $t_i$, inflation (or a similar mechanism) blow the initial quantum fluctuations and create a large homogeneous patch for our universe. In terms of our comoving coordinates, its size  is $\chi_\calC = c/ (a(t_i) H(t_i))$, where $H(t_i)^2 = 8\pi /3 G \rho(t_i)$ is given by the (potential) energy of inflation $\rho(t_i) \sim V(\phi)$.
 This comoving scale has remain constant and outside causal contact through out the evolution of the Universe. 
As we dont know the values of $a_i$ or $\rho(t_i)$ it seems impossible to estimate how large  $\chi_\calC$ is from current observations or first principles.
 
 But imagine that DE does not exist. Then, the horizon of our expanding universe (that emerged after inflation ended) will  eventually reach $\chi_\calC$ at some time $a_\calC$. This is illustrated in Fig.\ref{Fig:horizon}.
  If we assume that vacuum energy does not evolve after inflation (i.e. $\omega_{\rm vac}=-1$), we can use Eq.\ref{eq:rhoH}-\ref{eq:rhoH2} to estimate:
\beq
\rho_{\Lambda} = \rho_{\calC} =
 \frac{\int_{M_{\calC}}  \sqrt{-g} d^4x  (\rho_m + 2\rho_r) ~ }
 {\int_{M_{\calC}}  \sqrt{-g} d^4x }.
\label{eq:rhoH4}
\eeq
We can actually only do this calculation at time $a_\calC$ because (from current observations) we only know the full content of the Universe at that time. At earlier times, part of the causal region is outside our horizon. Thus, we can use our measurements
of $\rho_\Lambda$, $\rho_m$ and $\rho_r$ to estimate $a_\calC$ and therefore $\chi_\calC$.

The horizon after inflation
(see Eq.\ref{eq:eta}) is:
\beq
\chi(a) = \eta(a) - \eta(a_e)
\label{eq:chia}
\eeq
where $a_e$ represents the end of inflation. We then have $\chi_\calC = \chi(a_\calC) = \eta(a_i)$
where $a_{\calC}$ is the time when the causal boundary enters the horizon after inflation and $a_i$ the begining of inflation.  Fig.\ref{Fig:horizon} illustrate this.
We calculate $\rho_{\calC}$  in Eq.\ref{eq:rhoH4} as the integral to $\chi_{\calC}$ in the light-cone:

\beq
\rho_{\calC}  = 
~ \frac{\int_{0}^{\chi_{\calC}}  d\chi ~ \chi^2 ~{a^3}~(\rho_m a^{-3} + 2\rho_r a^{-4})~ }
{2 \int_{0}^{\chi_{\calC}}  d\chi  ~\chi^2 ~a^3 } ,
\label{Eq:rhoHchi}
\eeq
where $a=a(\chi)$ in Eq.\ref{eq:chia} and Eq.\ref{eq:eta}.
 For $H(a)$ we use Eq.\ref{eq:Hubble}
with $\Omega \equiv \rho/\rho_c$,   $\rho_c=  3H_0^2 / 8\pi G$ and
$\Omega_r=4.2 \times 10^{-5}$ \citep{Planck2018}
for a flat Universe
$\Omega_m=1-\Omega_\Lambda-\Omega_r$. 
We find $\chi_\calC$ from Eq.\ref{Eq:rhoHchi} numerically using $\Omega_\Lambda=\Omega_{\calC} \equiv \rho_\calC/\rho_c
\simeq 0.69 \pm 0.01$:

\bea
\chi_{\calC} &=& \left( 3.149 \pm 0.006 \right) \frac{c}{H_0} 
\label{eq:chi_H} \\
a_{\calC} &=& 0.933 \pm 0.006 .
\label{eq:a_H}
\eea
to be compared to $a_0=1$  and $\chi_0=3.200
\frac{c}{H_0}$ today. So we can see that the scale of our causal universe is slightly smaller than our observable universe today.
Because $\chi_\calC$  is smaller than $2\pi$ times our observable horizon, 
we should be able to see this horizon in our past lightcone
at  $\theta_\calC(z)=\chi_\calC/\chi(z)$. At $z\simeq 1$  about half of the sky ($\theta_\calC \sim 180$ deg) is causally disconnected. 
At larger redshifts this boundary tends to a fix value $\theta_\calC \simeqß 60$ deg. depending on  $\chi_\calC$ (and therefore $\Lambda$).
This has implications for CMB observations (see  section \ref{sec:CMB}).
{\color{black}
In Appendix \ref{sec:DE} we discuss how this results change in the presence of DE. But our proposal is that DE is not needed to explain cosmic acceleration.
 If we set $\Omega_r=0$ we find $a_{\calC}=0.86$ and  $\chi_{\calC}=3.081$, so our results are not very sensitive to the details of the Early Universe after inflation.
}

%The scale factor $a_\calC$ in Eq.\ref{eq:a_H} corresponds to an age: $t_{\calC}= \left( 0.795 \pm 0.004 \right) \frac{1}{H_0} \simeq 11.9$ Gyr , compared to $t_{age} \simeq 0.955/H_0$ today, i.e. about $\Delta t \simeq 2.4$ Gyr into our past.

%%%%%%%%%%%%%%%%%%%%%%%%%%%%%%%%%%%%%%%%%%
%\subsection{Time dependence and the coincidence problem}
\subsection{Inflation and the coincidence problem}
\label{sec:coincidence}

Do the results in previous section, e.g. Eq.\ref{eq:chi_H}, depend on the observer?
An astronomer in a galaxy at z=9, when $a =0.1$, will measure $\Omega_m$ to be
$\Omega_m(z=9)= \Omega_m(z=0) a^{-3} H_0^2/H^2 \simeq 0.997$ and $\Omega_\Lambda \simeq 0.002$. If she could measure these values with the accuracy that we measure then today she will get the exact same result for  Eq.\ref{eq:chi_H}-\ref{eq:a_H} as we do today. She will conclude that $\chi_\calC$ is larger than the observable universe at $z=9$ and will predict that that cosmic acceleration will happen in the future when the horizon reaches $\chi_\calC$.
So in that respect, there is no coincide problem in these results. They do not depend on when we estimate them.

\begin{figure}
\centering \includegraphics[width=\linewidth]{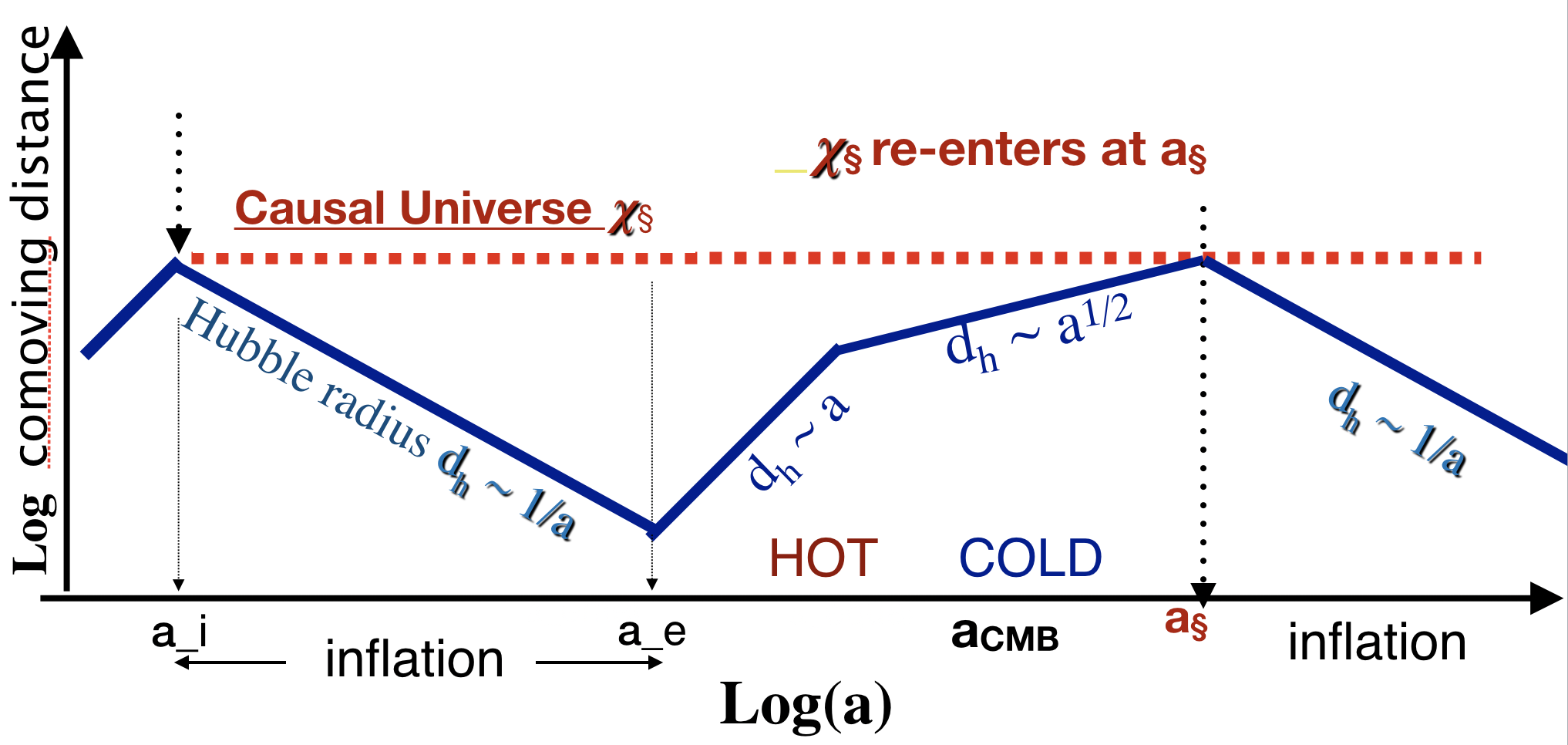}
\caption{
Same as Fig.\ref{Fig:horizon0}, but here the causal universe $\chi_\calC$ (set by inflation) is responsible for the late time cosmic acceleration.}
\label{Fig:horizon}
\end{figure}

Eq.\ref{eq:rhoH4} indicates that when the causal boundary is close to re-entering the Horizon the expansion becomes dominated by $\rho_\Lambda$. This is because $\rho_m(a_\calC)<\rho_\calC=\rho_\Lambda$, as density decreases with the expansion. This results in another inflationary epoch at $a=a_\calC$ which keeps the Causal Universe frozen (see Fig.\ref{Fig:horizon}).
We are trap inside the causal scale $\chi_\calC$ and causality can only play a role for comoving scales $\chi<\chi_\calC$. This warrants the constancy of $\Lambda$ and also explains cosmic acceleration as a consequence of (the first) inflation. 

 We can now recast the coincidence problem (why $\rho_{\Lambda} \simeq 2.3 \rho_m$?)
into a new question: why we live at a time which is close to  $a_\calC$? 
or why the scale/energy of inflation is close to our horizon/energy today and not larger or smaller?
In terms of anthropic reasoning  \citep{Weinberg1989,2003PhRvD..67d3503G}, 
at earlier times the Universe is dominated by radiation and there are no stars or galaxies to host observers. 
Closer to $a\simeq a_\calC$ the Universe is dominated by matter and there are galaxies and stars with planets and potential observers.
At later times $\Omega_\Lambda \simeq 1$ and galaxies will be torn apart by the new inflation. 
Moreover, $a_\calC$ has the largest Hubble radius (see Fig.\ref{Fig:horizon}) with the highest chances to host observers like us.
There is nothing too special about this coincidence.
Ultimately, the reason why $\chi_\calC \sim 3c/H_0$  reside in the details of inflation: when inflation begins $a_i$ and ends $a_e$ (see Fig.\ref{Fig:horizon}).
This recasts the coincidence problem into an opportunity to better understand inflation and the origin of homogeneity. We propose here to identify $\chi_\calC=\eta(a_i)$ with the comoving  horizon before inflation begins at time $t_i$:

\begin{equation}
a_i H_i =  c\chi_{\calC}^{-1} \simeq \left( 0.3176 \pm 0.0006\right) H_0
%a_i H_I =  c\chi_{\calC}^{-1} \simeq \left( 0.2940 \pm 0.004\right) H_0
\label{eq:aH}
\end{equation}
where $H_i=H(t_i)$ or $a_i=a(t_i)$.
 This shows how $a_i$ determines $\chi_\calC$, while $a_e$ determines when it re-enters the horizon (see Fig.\ref{Fig:horizon}), and therefore how large  $\rho_\calC$ is.
The Hubble rate during inflation $H_I$ is proportional to the energy of inflation. During reheating this energy is converted into radiation:
$H_I^2 \simeq \Omega_r ~H_0^2 ~ a_e^{-4}$, with
$a_e \equiv e^{N} a_i$. We can combine  with Eq.\ref{eq:aH} to find:

\beq
a_i \chi_{\calC} = \frac{H_i}{H_I} e^{-2N} ~ \Omega_r^{1/2} ~ (\chi_{\calC}^2 H_0/c) \simeq 4 \times 10^8
~ l_{\rm Planck} 
\label{eq:achi}
\eeq
where for the second equality we have used the canonical value of $N \simeq 60$
and $H_i \simeq H_I$, which also yields $a_i \simeq 1.56 \times 10^{-53}$
and $H_I \sim 10^{10}$ GeV.
 The condition  $a_i \chi_{\calC}> l_{\rm Planck}$ requires $N<70$, close to the value found in \cite{DodelsonHui2003}. 
Thus, the whole causal size of our Universe $\chi_\calC$ could result from a quantum  fluctuation at the Planck scale $l_{\rm Planck}$. Such vacuum fluctuation could generate an inflationary expansion. 
After $N \simeq 70$ e-folds inflation ends with
reheating, which results into matter and radiation today. Thus, this model links the cosmological constant scale to the Planck scale via inflation.

%%%%%%%%%%%%%%%%%%%%%%%%%%%%%%%%%%%%%%%%%%
\subsection{Implications for CMB}
\label{sec:CMB}

\begin{figure}
\vspace{-1.5cm}
\includegraphics[width=\linewidth]{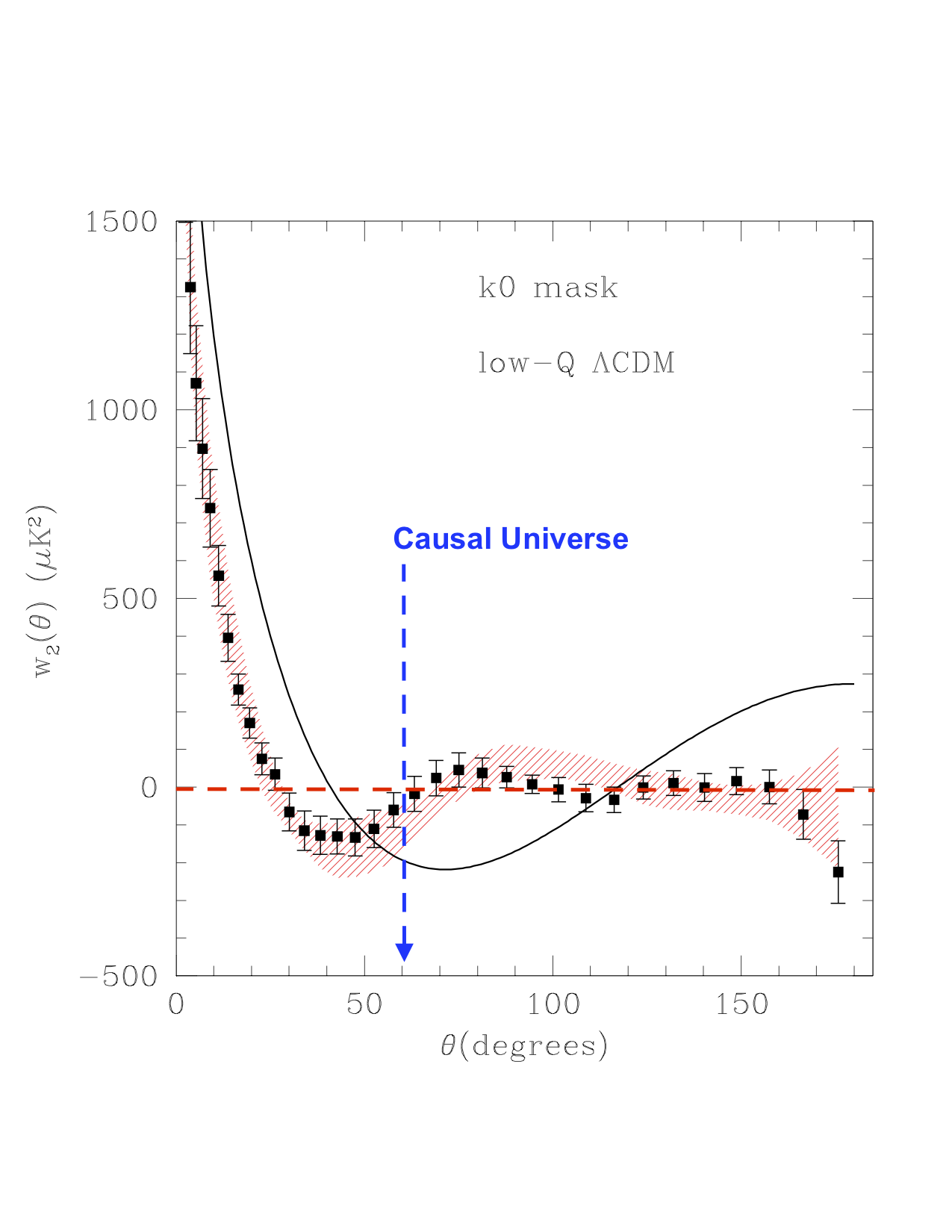}
\vspace{-1.5cm}
\caption{
Two-point correlation function of measured CMB temperature fluctuations in WMAP (points with errorbars) as a function of angular separation  (from \citealt{Gaztanaga2003}). The black continuous line shows the  $\Lambda$CDM prediction for an infinite Universe. 
Shaded region shows  $\Lambda$CDM simulations where 
we suppress the large scale modes (multipoles $l<5$).
}
\label{Fig:CMB}
\end{figure}

The (look-back) comoving distance to the surface of last scattering $a_* \simeq 9.2 \times 10^{-4}$ \citep{Planck2018} is
$\chi_{CMB} = \eta(1)- \eta(a_*) \simeq 3.145 ~ \frac{c}{H_0}$. This is similar to our estimate for $\chi_\calC$ in Eq.\ref{eq:chi_H}.
Thus, we would expect to see no correlations in the CMB on angular scales  $\theta > \theta_\calC \equiv \frac{\chi_\calC}{\chi_{CMB}} \simeq 60$ degrees 
for $\Omega_\Lambda =\Omega_\calC \simeq 0.7$.
The lack of structure seen in the CMB on these large scales is one of the well known anomalies in the CMB data, see \cite{Schwarz2016} and references therein. 
{\color{black}
This lack of correlations above 60 deg. has been interpreted as a universe with non trivial topology \citep{Luminet2003}.}
Fig.\ref{Fig:CMB} shows a comparison of the measured CMB temperature correlations (points with error-bars) with the $\Lambda$CDM prediction  for an infinite Universe (continuous line). There is a very clear discrepancy
which \cite{Copi2009} estimates to happen in only 0.025 per cent of the realizations of the infinite $\Lambda$CDM model. 
The significance of this discrepancy is model dependent. If errors are estimated from the data (and not from the model)
$\Lambda$CDM is strongly ruled out by this measurement \citep{Gaztanaga2003}.
Even assuming a $\Lambda$CDM model, the lack of large scale correlations in $w_2(\theta)$ represent an odd alignment of lower order multipoles of the angular power spectrum $c_l$ \citep{Schwarz2016}.

A small causal universe also predicts similar lack of correlation for CMB polarization measurements. We also expect variations of $c_l$ on smaller scales (e.g. around the first acoustic peak $l \simeq 200$) for $c_l$ estimated over different regions of the sky (separated by $\theta>60$ deg.), which is another known CMB anomaly \citep{PlanckCl-asym}. Early evidence for these variations \citep{gazta1998} were interpreted as non-Gaussian initial conditions.

We can also predict $\Omega_\Lambda$ from the lack of CMB correlations. From Fig.\ref{Fig:CMB} we roughly estimate $\theta_\calC \simeq 60 \pm 3$ deg. to find (using Eq.\ref{Eq:rhoHchi}) $\Omega_\Lambda = 0.7 \pm 0.1$. 
(the larger the angle the smaller $\Omega_\Lambda$).
But note that this rough estimate does not take into account the foreground (late) ISW and lensing effects \citep{Fosalba,ISW}, which add non primordial correlations to the largest scales.
This requires further investigation. Also note  that this estimate for $\Omega_\Lambda$
corresponds to the size of disconnected regions at the location of the CMB, which might be slightly different to the value near us, as we see a different patch of the primordial Universe (see bellow).
Note also that there are temperature differences on scales larger $\theta_\calC$, but they are not correlated, as expected in causality disconnected regions.
Nearby regions are connected which creates a smooth transition across disconnected regions.

%%%%%%%%%%%%%%%%%%%%%%%%%%%%%%%%%%%%%%%%%% 
\section{Discussion and Conclusions}

$\Lambda$CDM in Eq.\ref{eq:frw}-\ref{eq:Hubble}  assumes that $\rho$ is constant everywhere at a fixed comoving time.
This requires acausal initial conditions \citep{Brandenberger} unless there is inflation, where a tiny homogeneous and causally connected patch, the Causal Universe $\chi_\calC$, was inflated to be very large today. Regions larger than  $\chi_\calC$ are out of causal contact. Here we require that 
test particles become free (or the relativistic flux is zero)
as we approach $\chi_\calC$. No cause should produce no effect.
This leads to Eq.\ref{eq:rhoH}, which is the main result in this paper.
If we ignore  the vacuum, this condition requires:
$\Lambda  = 8\pi G \rho_{\calC}$, where $\rho_\calC$ is the  matter and radiation inside 
$\chi_\calC$ (Eq.\ref{eq:rhoH4}). For an infinite Universe
($\chi_{\calC} \rightarrow \infty$) we have $\rho_{\calC} \Rightarrow 0$ which requires $\Lambda \Rightarrow 0$. This is also what we find in classical gravity with a $\Lambda$ term, because Hooke's term diverges at infinity (see Eq.\ref{eq:hooke}).  So  the  fact that $\Lambda \neq 0$   could indicate that  $\chi_\calC$ is not  infinite. Adding vacuum  $\rho_{\rm vac}$ does not change this argument because  $\rho_\Lambda \equiv \Lambda/ 8\pi G + \rho_{\rm vac} = \rho_{\calC}$ turns out to be independent of $\rho_{\rm vac}$
(see Eq.\ref{eq:rhoH2}). 
Thus, whether the causal size of the Universe is finite or not, $\rho_{\rm vac}$ can not gravitate! The cancellation between $\Lambda$ and $\rho_{\rm vac}$ is a direct consequence of the boundary condition and it also implies that deSitter Universe (empty with a cosmological constant) is not causal (it produces curvature even when empty) and therefore not a physical model.
%In classical terms of Eq.\ref{eq:hooke}, $\Lambda$ will produce a  divergent gravitational force.

For constant vacuum ($\omega \equiv p/\rho =-1$), we find  $\chi_\calC \simeq 3 c/H_0$ 
for  $\Omega_\calC= \Omega_\Lambda \simeq 0.7$. 
We can also estimate $\chi_\calC$ as $c/(a_i H_i)$ 
when inflation begins, see Eq.\ref{eq:aH}. After inflation  $\chi_\calC$ freezes out until it re-enters causality at $a_\calC \simeq 0.93$, close to now ($a=1$). This starts a new inflation (as $\rho_\Lambda = \rho_{\calC}> \rho_m$) which keeps the causal boundary frozen. 
Thus a finite  $\chi_\calC$  explains why $\rho_\Lambda \simeq 2\rho_m$.
It also predicts that CMB temperature should not be correlated above $\theta> \theta_\calC \simeq 60$ deg. A prediction that matches observations
(see Fig.\ref{Fig:CMB}). This is one of the well known anomalies measured in the CMB.
One would also expect the CMB spectrum to be anisotropic on the largest scales, which is another well known measured anomaly (see \citealt{PlanckCl-asym}).
One can  reverse this argument to use the lack of CMB correlations above
$\theta_\calC \simeq 60$ deg, to estimate $\chi_\calC \simeq \theta_\calC \chi_{CMB}$. Together with condition $\rho_\calC=\rho_\Lambda$, this  provides a  prediction of $\Omega_\Lambda \simeq 0.7 \pm 0.1$, which is independent of other measurements for $\Omega_\Lambda$.
 More work is needed to account for the late ISW and lensing  
 and to interpret the CMB measurements with a metric that is not homogeneous \citep{Gaztanaga2019}.

Note that because the Universe is not strictly homogeneous outside a causal region, the causal boundary for observers far away from us could be slightly different from ours, because they see a different patch of the Universe which could have slightly different energy content. Continuity across nearby disconnected regions forces these differences to be small, but it is impossible to quantify this without a model for the initial conditions and a better understanding of the process that generates the primordial homogeneity. 
In general such differences could affect structure formation, galaxy evolution and CMB observations. The fact that we can measure a concordance picture from different observations with the $\Lambda$CDM model indicates that these differences must be small. But tensions between measurements of cosmological parameters (or fundamental constants of nature) from very different redshifts (eg between CMB and local measuremnts) or different parts of the sky at high redhifts (such as dipolar variation of fine structure constant  in \citealt{Webb2011}) could be related to such in-homogeneities, rather than to evolution of the Dark Energy (DE) equation of state or other more exotic explanations.

For $\chi_\calC>> 3c/H_0$  we can not explain cosmic acceleration with $\omega=-1$, because  the resulting $\rho_\Lambda$  in Eq.\ref{eq:rhoH4} would be very small. We need evolving DE with  equation of state $\omega_{DE}>-1$ and $\rho_\Lambda = \rho_{DE}$ today (see Appendix \ref{sec:DE}).  But DE gives no clue as to why  $\rho_{DE} \simeq 2 \rho_{m}$ today and can not explain  the anomalous lack of CMB correlations at large scales. 
We can apply Occam's razor to argue that there is no need for DE or Modify Gravity. Comparing Fig.\ref{Fig:horizon0} with  Fig.\ref{Fig:horizon}, we can see that the measured cosmic acceleration today can be explained by a finite causal scale 
$\chi_\calC \simeq 3 c/H_0$, which is expected for Universe with a finite age and a mechanism like inflation.
 
We have speculated that the finite causal size of our universe 
could result from a quantum fluctuation at the Planck scale $l_{\rm Planck}$ which produces 
an inflated expansion by a factor $\simeq e^{70}$ 
leading to a reheating into matter and radiation today. 
{\color{black} {}This eventually results into late time cosmic acceleration.
%, a measurement that can then be interpreted as the smoking gun of primordial Inflation.
The existence of such finite causal scale can be tested  further with observations of the variation of cosmological or fundamental parameters over cosmological scales.}

\section*{acknowledgments}
I want to thank A.Alarcon, J.Barrow, C.Baugh, R.Brandenberger, G.Bernstein, M.Bruni, S.Dodelson,  E.Elizalde,  P.Fosalba, J.Frieman, M.Gatti, L.Hui, D.Huterer, A. Liddle, M.Manera, P.J.E. Peebles, R.Scoccimarro, P.Renard, I.Tutusaus and S.Weinberg for their feedback.
This work has been supported by MINECO  grants AYA2015-71825, PGC2019-102021-B-100, LACEGAL Marie Sklodowska-Curie grant No 734374 with ERDF funds from the EU  Horizon 2020 Programme. IEEC is partially funded by the CERCA program of the Generalitat de Catalunya. 

\bibliographystyle{mnras}
\bibliography{gaztanaga} % if your bibtex file is called example.bib

\appendix

\section{Hooke's law}
\label{sec:hook}

In Newton's gravity, a point mass $m$ generates a radial acceleration
$\vec{g} = - \vec{\nabla} \phi  = -G m \vec{r}/r^3$, 
where $\phi$ is the Newtonian potential.
The $\Lambda$ term in GR's field equations, Eq.\ref{eq:rmunu},  corresponds here to an additional term which is linear in $r$, as in Hooke's law:

\beq
\vec{g} =  - \vec{\nabla} \phi  = \vec{g}_{Newton} + \vec{g}_{Hooke} =
- \left(
\frac{G}{r^3} + G_2   \right)~ m ~  \vec{r}
\label{Hooke}
\eeq
Hooke's constant, $G_2$, can be related to $\Lambda$, as we will see below.
This Hooke's term is unique in that it is the only distance dependence (other than the inverse square law)  that has a key property for gravity: that a spherical mass shell of arbitrary density $\rho(R)$ produces a gravitational field which is identical to a point source of equal mass in its center:

\beq
\vec{g}_{Hooke} = - G_2 \int_{shell} d^3R ~  (\vec{r} - \vec{R}) ~\rho(R)= - G_2 m ~\vec{r}
\eeq
where $ \vec{R}$ covers the shell and $\vec{r}$ is some position outside. This is a key property of gravity, needed to treat the Universe as a whole
and sustain  Gauss's and Birkhoff's theorems.
Thus in Classical Mechanics the above law of gravity Eq.(\ref{Hooke}) is consistent
with the symmetries and observations, as long as $G_2$ is small
enough. 
In fact Newton, and other scientist, had already noticed this, but did not consider
Hooke's term for lack of observational evidence \cite[and references therein]{CalderLahav2008}. 

Notice how Eq.\ref{Hooke} diverges for $r \Rightarrow \infty$. We expect instead that $\vec{g} \Rightarrow 0$, as particles should be free at infinity (as there could be no causal connection). This explains why on theoretical grounds we would expect $G_2=0$ even if this relation is allowed by the symmetries of the problem. On the other hand, if causality ends at some finite distance $r_{\calC}$, as the Universe has a finite age, then we see that Newton's law alone can not describe a causal gravitational interaction. We need to also have Hooke's law.
 Requiring that gravitational forces $\vec{g}$ in Eq.\ref{Hooke} vanished at the causal boundary $r_{\calC}$:

\beq
\vec{g}(r=r_{\calC}) = 0  ~~ \Rightarrow ~~   G_2 = - 
\frac{4\pi G}{3 V_{\calC}} 
\label{eq:g=0}
\eeq
so that Hooke's gravitational force is repulsive and only becomes comparable to Newton's gravity for separations comparable to  $r_{\calC}$.

We can  estimate the flux  using Eq.(\ref{Hooke}) to find:

\beq
\Phi= 
\oint_{\partial V} ~d\vec{\mathrm{r}} ~ \vec{\mathrm{g}}  = 
 -4\pi G m_V - 3 V G_2 m_{\calC}
\label{eq:fluxHook}
\eeq
where $m_V$ is the mass inside $V$ and
$m_{\calC}$ is all the mass in all the (causally connected) universe. As expected,  $\Phi=0$ in Eq.\ref{eq:fluxHook} reproduces Eq.\ref{eq:g=0} for $V=V_\calC$.
For the Newtonian term,
masses outside $V$ do not contribute to $\Phi$ because for a small cone center outside and crossing $V$ the in going flux exactly cancels with the outgoing flux (as the inverse square law in $\vec{g}$ compensates the increase in the area crossed then the cone goes inside to when is going outside). This is not the case for the Hooke's law, where the difference in flux is just  proportional to the volume inside $V$. 
As all the masses contribute equally, we have:

\beq
\Phi_{Hooke}(V) = - G_2   \sum_m ~m(\vec{r}) ~\oint_{\partial V} ~(\vec{R} - \vec{r}) ~d\vec{A} = - 3 V G_2 m_{\calC}
\label{eq:Hookeflux}
\eeq
where the total mass is $ m_{\calC}= \sum_m  m$.
This result is also true in arbitrary number of dimensions (see \cite{Wilkins})). 
This explains why Hooke's law behaves like vacuum energy (i.e. like negative pressure: $\omega=-1$) or $\Lambda$ in GR: the density
remains constant as we increase the volume!
Comparing Eq.\ref{eq:fluxHook} with  Eq.\ref{eq:flux00} we have:
\beq
\Lambda  =  -3 G_2 m_{\calC} = 4\pi G \rho_\calC
\label{eq:G2}
\eeq
where in the last step we have also used Eq.\ref{eq:g=0}, or equivalently $\Phi=0$ in  Eq.\ref{eq:fluxHook}. This of course agrees with Eq.\ref{eq:lambda0}, which shows that 
Eq.\ref{Hooke} and Eq.\ref{eq:hooke} are equivalent.

This result provides a Classical Physic's interpretation of the 
cosmological constant: it is just related to Hooke's constant $G_2$ and the total mass $m_{\calC}$ in the (causal) universe. Note how $\Lambda$ can be zero if either $G_2=0$ or if $m_{\calC}=0$.  Note also how $G_2=0$ for infinite volume and in this case $\Lambda=0$ because $m_{\calC}/V_{\calC}$ also goes to zero. This relation establishes a clear connection between the value of $\Lambda$ and the matter-energy content in the Universe, as requiered by causality. It also gives new light into Mach's Principle.

{\color{black}
%%%%%%%%%%%%%%%%%%%%%%%%%%%%%%%%%%%%%%%%%%
\section{Effective Dark Energy (DE)}
\label{sec:DE}
%\medskip

Here we generalize the results of section \ref{sec:size} to the case with DE.
If vacuum energy suffers a phase transition or changes with time, as could have happened during inflation, the cancellation 
presented in section \ref{sec:vac}
will not happen (because $\omega \neq -1$) and an evolving  $\rho_{\rm vac}$ (which we usually call Dark Energy) could contribute to the effective value of $\rho_{\Lambda}$. Consider the general case of DE after inflation:
\bea
\rho_{DE}(a)  &=& \rho_{\rm vac} + \rho_{DE} ~a^{-3(1+\omega)}
\label{eq:DE}
\\ \nonumber
p_{DE}(a)  &=& -\rho_{\rm vac} + \omega ~\rho_{DE} ~ a^{-3(1+\omega)} ,
\eea
where (by definition) only one component of DE is evolving.
%We assume a constant value for $\omega$, but the analysis can easily be extended to evolving values of $\omega$.
We  then have from Eq.\ref{eq:rhoH} and Eq.\ref{eq:rhoHlambda}:
\beq
\rho_{\Lambda} 
%= {\Lambda \over{8\pi G}} + \rho_{\calC} +  \rho_{\rm vac} + \rho_{DE}
= {\rho}_{\calC} +
 \rho_{DE} [1 +  \frac{1+3\omega}{2} \hat{a}_{\calC}^{-3(1+\omega)}] .
 \label{eq:rhoH3b}
\eeq
where $\hat{a}_{\calC}$ is some mean value of $a$ in the past light-cone of $a_\calC$ in Eq.\ref{Eq:rhoHchi}.
This reduces to  $\rho_\Lambda = \rho_\calC$ for $\omega=-1$.
For $a_\calC \Rightarrow \infty$ 
we have $\rho_\calC \Rightarrow 0$  because  $\rho_m(a)$ and $\rho_r(a)$ tend to zero as we increase $a_\calC$. The same happens with $\hat{a}_{\calC}^{-3(1+\omega)}$ for $\omega >-1$, so that:
\beq
\rho_\Lambda  = \rho_{DE}   ~~ \rm{for}~~ {a}_\calC \Rightarrow \infty  ~~\rm{\&} ~~ \omega>-1 .
\label{eq:rhoDE}
\eeq
% For $\omega<-1$ we have $\rho_\Lambda \Rightarrow \infty$. 
 %But $\omega > -1$  will work even for values of $\omega$ which are arbitrarily close to $-1$.
So evolving DE  could produce the observed cosmic acceleration in an infinitely large Universe. 
This solution does not explain why $\rho_{\Lambda} = \rho_{DE} \simeq 2\rho_m$. The original motivation to introduce DE was to explain how vacuum energy $\rho_{\rm vac}$ could be as small as the measured $\rho_\Lambda$ \citep{Weinberg1989,Huterer}. But we have shown in Section \ref{sec:vac} that the causal boundary condition explains why $\rho_{\rm vac}$ does not contribute to $\rho_\Lambda$ and also results in $\rho_{\Lambda} \simeq 2\rho_m$. This removes the motivation to have DE, as it represents an unnecessary complication of the model (Occam's razor).

If observations find $\omega$ to be significantly larger than $\omega =-1$ this will indicate that something like DE exist. The 
actual measured value of $\omega$ will not directly tell us the size of $\chi_\calC$ unless we also have some model for $\rho_{DE}$. But very accurate measurements for the evolution of $\rho_\Lambda$ might actually be able to separate a component that is constant (i.e. $\rho_\calC$) from a component that is evolving (i.e. DE).

}

% Don't change these lines
\bsp	% typesetting comment
\label{lastpage}
\end{document}